\documentclass[12pt,letterpaper]{article}

\newif\ifarxiv
\arxivtrue

\usepackage[margin=1in]{geometry}
\usepackage{iftex}
\ifPDFTeX
  \usepackage[T1]{fontenc}
  \usepackage[utf8]{inputenc}
  \usepackage{mathptmx}
\else
  \usepackage{fontspec}
\fi
\usepackage{setspace}
\usepackage{amsmath}
\usepackage{amssymb}
\usepackage{ragged2e}
\usepackage{indentfirst}
\usepackage{graphicx}
\usepackage{booktabs}
\usepackage{multirow}
\usepackage{array}
\usepackage{tabularx}
\usepackage{caption}
\usepackage{float}
\usepackage{pdflscape}
\usepackage{enumitem}
\usepackage{fancyhdr}
\usepackage[hidelinks]{hyperref}
\usepackage{color}
\usepackage{xr}
\usepackage{subfiles}
\ifarxiv
\else
  \makeatletter
  \newcommand*{\addFileDependency}[1]{%
    \typeout{(#1)}%
    \@addtofilelist{#1}%
    \IfFileExists{#1}{}{\typeout{No file #1.}}%
  }
  \makeatother
  \addFileDependency{web_appendix.tex}%
  \addFileDependency{web_appendix.aux}%
\fi

\RaggedRight
\setlist{itemsep=0pt, parsep=0pt, topsep=0pt}
\newcolumntype{Y}{>{\RaggedRight\arraybackslash}X}
\newenvironment{tabitemize}{%
  \begin{itemize}[leftmargin=*, nosep, topsep=0pt, partopsep=0pt, parsep=0pt, itemsep=0pt]%
}{%
  \end{itemize}%
}

\newcommand{\primaryheading}[1]{%
  \par\vspace{\baselineskip}%
  \begin{center}\textbf{\MakeUppercase{#1}}\end{center}%
  \vspace{\baselineskip}%
}

\newcommand{\secondaryheading}[1]{%
  \par\addvspace{\baselineskip}%
  \noindent#1\par\nobreak
  \addvspace{\baselineskip}%
}

\newcommand{\tertiaryheading}[1]{%
  \par\addvspace{\baselineskip}%
  \noindent\hspace*{0.5in}\textit{#1.}\ %
}

\graphicspath{{figures/}}

\newcounter{webappendix}
\renewcommand{\thewebappendix}{\Alph{webappendix}}

\newcommand{\webappendixheading}[2]{%
  \clearpage
  \refstepcounter{webappendix}\label{#2}%
  \setcounter{table}{0}%
  \setcounter{figure}{0}%
  \renewcommand{\thetable}{\thewebappendix\arabic{table}}%
  \renewcommand{\thefigure}{\thewebappendix\arabic{figure}}%
  \begin{center}
    \textbf{WEB APPENDIX \thewebappendix}\\[\baselineskip]
    \textbf{\MakeUppercase{#1}}
  \end{center}
  \vspace{\baselineskip}
}

\newcommand{\articletitle}{%
  Brief Commentary: ExploraTwin, a Non-Profit Research Platform for Digital Twin Simulations
}

\usepackage{citation-style-language}

\cslsetup{
  style = journal-of-consumer-research,
  bib-font = {\doublespacing\RaggedRight},
  bib-hang = 0.5in,
  bib-item-sep = 0pt
}

\begin{document}

\begin{center}
\textbf{ExploraTwin, a Non-Profit Research Platform for Digital Twin Simulations}
\vspace{2em}

\textbf{Naveen Venkat}\\
Columbia University\\
\href{mailto:nv2444@columbia.edu}{nv2444@columbia.edu}

\vspace{1.5em}

\textbf{Yuchen Qiu}\\
Columbia Business School\\
\href{mailto:yq2411@columbia.edu}{yq2411@columbia.edu}

\vspace{1.5em}

\textbf{Tianyi Peng}\\
Columbia Business School\\
\href{mailto:tianyi.peng@columbia.edu}{tianyi.peng@columbia.edu}

\vspace{1.5em}

\textbf{George Gui}\\
Columbia Business School\\
\href{mailto:zg2467@gsb.columbia.edu}{zg2467@gsb.columbia.edu}

\vspace{1.5em}

\textbf{Olivier Toubia}\\
Columbia Business School\\
\href{mailto:ot2107@gsb.columbia.edu}{ot2107@gsb.columbia.edu}

\end{center}

 \newpage

\primaryheading{Abstract}
Digital twin simulations show promise, but current empirical evidence suggests that the approach should be tested before being deployed in any particular context. To lower the friction for researchers and practitioners to test and deploy digital twin simulations, this brief commentary introduces \href{https://exploratwin.org}{ExploraTwin}, an open-access, non-profit research platform for digital twin survey simulations. ExploraTwin supports two modes. In survey mode, researchers can upload a Qualtrics survey file or create a survey within the platform; select an available sample of digital twins; configure and run the simulation, and export analysis-ready data. In panel mode, researchers can assemble a small group of twins for open-ended conversations, document annotation, and moderated, focus-group-style voice discussions. We also developed CroissantTwin, a standardized data format for adding samples of digital twins to the platform. We demonstrate the survey mode workflow by using the platform to replicate 19 experiments on digital twins from the Twin-2K-500 dataset. ExploraTwin's survey execution fidelity is high: 99.6\% of 197,000 answer units returned a structurally valid response on the first run. 


\clearpage

Large language models (LLMs) offer researchers the \emph{potential} to field a survey to an entire panel of simulated respondents almost instantaneously, at a small fraction of the cost of human data collection, using instruments and stimuli that have never been shown to anyone \citep{AherEtAl2023SimulateMultipleHumans, ArgyleEtAl2023OutOfOneMany, ParkEtAl2023GenerativeAgents,PengEtAl2025FunhouseMirrors, ToubiaEtAl2025Twin2K500,ManningHorton2026GeneralSocialAgents,AshokkumarEtAl2026SocialScienceExperiments}.

A particularly promising approach for doing so leverages digital twins, i.e., prompting LLMs with persona information---individuals' attributes, behavioral histories, survey responses, or open-ended reflections---to simulate specific human respondents in experiments \citep{ParkEtAl2024SelfReports,PengEtAl2025FunhouseMirrors,ToubiaEtAl2025Twin2K500}. For academic researchers and market research practitioners, digital twins potentially offer a fast and economical way to pretest survey materials, explore hypotheses, screen market research questions before costly human data collection, or revisit completed studies with additional questions. 

So far, the predictive performance of digital twins has been mixed \citep{ParkEtAl2024SelfReports, ToubiaEtAl2025Twin2K500,PengEtAl2025FunhouseMirrors}. This underscores the need for researchers and practitioners to experiment with and test digital twin pipelines before deploying them for a particular use case. This in turn requires tools that enable researchers to run studies on digital twins at low cost and with low friction. Despite the low marginal API cost per synthetic respondent, current solutions typically involve either subscribing to a commercial tool or developing one's own digital twin pipeline. The latter presents at least two types of friction.

First, simulation pipelines are costly to engineer because consumer research uses a wide range of instruments and potentially complex structures. Consider a Qualtrics survey as an example. It might contain a mix of open-ended, scale, multiple choice questions, as well as features like block randomization, branching, loop and merge, etc. Simulating such a study requires tools that can parse the instrument, execute or flag its logic, and return analysis-ready data. A shared workflow for adapting diverse study formats would substantially reduce the engineering burden on researchers.

Second, samples of digital twins, which we refer to as ``persona banks,'' are difficult to transport across studies and tools. A persona bank is a dataset containing a sample of synthetic personas. Over the last two years, several such datasets have been created. These datasets differ in population, structure, and form of the persona itself: prior survey answers with structured and unstructured attributes \citep{ToubiaEtAl2025Twin2K500}, interview-derived scripts \citep{ParkEtAl2024SelfReports}, behavioral logs \citep{WangEtAl2026OPeRA}, and synthetic profile collections and narratives \citep{GeEtAl2024PersonaHub,NVIDIA2026NemotronPersonas}. Without a common standard that tells the simulation tool how to understand them, researchers cannot easily use different persona datasets in the same simulation workflow. As a result, each new persona dataset requires a custom way to load and process it for use in a simulation.


In this brief commentary, we introduce \href{https://exploratwin.org}{ExploraTwin} (https://exploratwin.org), an open-access, non-profit research platform that implements a workflow for running digital twin survey simulations with minimal cost and friction. In terms of cost, ExploraTwin is currently free (with a monthly allowance of \$10 in usage credits). Future versions might charge nominally to cover API and development costs (on the order of \$0.01 per synthetic respondent). In terms of friction, the platform is able to handle a wide range of instruments, allowing users, for example, to simply upload a Qualtrics survey file (.qsf) and receive simulated data in the same format as with human respondents (.csv), in a matter of minutes. While the platform's primary persona bank is the Twin-2K-500 sample of digital twins \citep{ToubiaEtAl2025Twin2K500}, users can also choose from various other persona banks. CroissantTwin further provides a standardized data format for creating new persona banks (e.g., focused on specific regions or topics).  



In sum, ExploraTwin turns digital twin simulation from a heavy engineering problem into a standardized workflow that social science researchers and market research practitioners can inspect, reuse, and evaluate. We release an open-source package for ExploraTwin’s survey-mode pipeline on \href{https://github.com/nav-v/surveytwin-oss.git}{GitHub (https://github.com/nav-v/surveytwin-oss.git)} and also maintain a hosted non-profit web instance (\href{https://exploratwin.org}{https://exploratwin.org}).

The rest of this brief commentary proceeds as follows. First, we introduce the ExploraTwin workflow and walk through the two simulation modes (survey and panel). Second, we describe CroissantTwin, our protocol for adding new persona banks to the platform. Third, we examine survey fidelity by replicating the 19 Mega-study experiments conducted by  \citet{PengEtAl2025FunhouseMirrors}. Finally, we discuss cost efficiency, limitations, and directions for future research.

\primaryheading{The ExploraTwin Pipeline: Survey Mode}

ExploraTwin offers two ways to conduct studies with digital twins, distinguished by the form of response the researcher wants to produce. Survey mode works like fielding an online survey to a sample of respondents. Researchers can either (i) upload a Qualtrics survey file (.qsf), or (ii) compose a survey in an interactive builder, either starting from a blank canvas with drag-and-drop editing or from a draft that ExploraTwin generates from a typed research question. The panel mode, introduced in the next section, allows the user to assemble a small group of twins in a conversational interface for qualitative interviews, document annotation, and focus-group-style discussions. 

\begin{figure}[H]
\centering
\includegraphics[width=\linewidth]{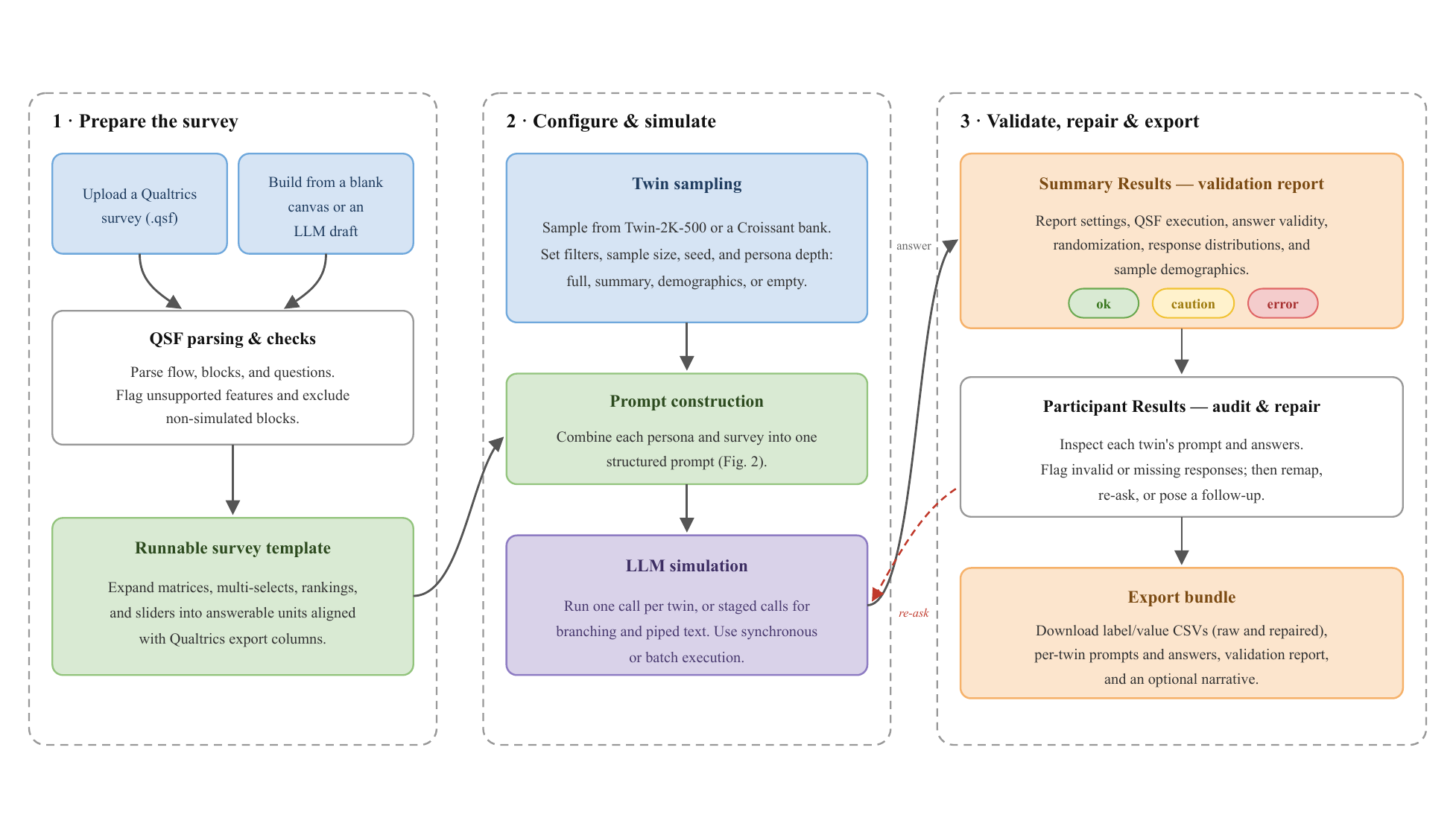}
\caption{SURVEY-MODE WORKFLOW FROM PREPARATION THROUGH VALIDATION AND EXPORT}
\label{fig:survey-pipeline}
\end{figure}

Figure~\ref{fig:survey-pipeline} summarizes the survey-mode workflow from researcher input through prompt construction, configuration, response generation, validation, repair, and export. Execution depends on the survey's logic. In static surveys, each twin completes the full survey in one model call: the questionnaire is presented once, and the model returns all answers in the requested format. If later questions depend on earlier answers (e.g., branching based on an earlier response), the run is staged: ExploraTwin first collects the answers needed to resolve the next part of the survey, applies the flow logic, and then presents the remaining questions with the twin's earlier answers. 

\secondaryheading{Preparing the Survey}
ExploraTwin provides two ways to prepare a survey to be fielded to the selected twins.

\tertiaryheading{From a QSF} The researcher uploads a survey in Qualtrics' native format, .qsf. The ExploraTwin parser then automatically converts the file into a runnable survey template. The parser supports a broad set of Qualtrics question types and logic features. See a complete list in table~\ref{tab:survey-feature-support}. If the uploaded survey contains features the pipeline cannot fully reproduce, such as custom Qualtrics JavaScript or externally defined embedded-data fields, ExploraTwin surfaces warnings before the run rather than silently translating them. The researcher can inspect the parsed block structure and drop blocks that should not be simulated, such as consent, debriefing, or administrative blocks. 

The parser also supports the inclusion of images in the survey. If the QSF contains visual stimuli, ExploraTwin automatically extracts them and pairs them with the surrounding text so vision-capable AI models can consider both when responding. 

\tertiaryheading{From the Survey Builder} Alternatively, the researcher can compose the survey in an interactive builder, starting either from a blank canvas or from a generated draft. In the latter case, the researcher types a research question or brief, and an LLM agent prompted as a survey designer drafts a structured questionnaire as a starting point. In both cases, the survey can include single-choice items, multi-select questions, rating scales, matrix items, open text, and bounded numeric-entry questions. And the researcher revises it in place by editing question wording, adding or removing items, attaching image stimuli, or requesting natural-language edits from the survey-design assistant. Once approved, the survey is converted into the same parsed-template structure used by the QSF pipeline. The builder targets straightforward linear instruments; complex survey logic---randomization, branching, display logic, piped text---requires the QSF path described above.

\secondaryheading{From Survey to Simulation Prompt}

An effective simulation is not just pasting survey questions into a prompt. Human respondents experience a survey as an interactive instrument, with the interface controlling question order, branching, display rules, and allowed response formats. By contrast, an LLM reads the survey as a token sequence. Therefore, the survey has to be translated into an LLM-readable template that preserves this survey-taking logic while specifying how the model should view and answer each question. For example, a human respondent answers a slider by dragging a control between labeled endpoints, whereas the LLM receives the question and endpoint labels as text, together with the permitted numeric range and an instruction to return a single value in the required format. The description of the selected persona is then combined with this template, and the model is instructed to return choices or text in a structured format. This translation layer lets the simulation approximate the logic of human survey-taking while working within the input and output constraints of LLMs.

\begin{figure}[H]
\centering
\includegraphics[width=\linewidth]{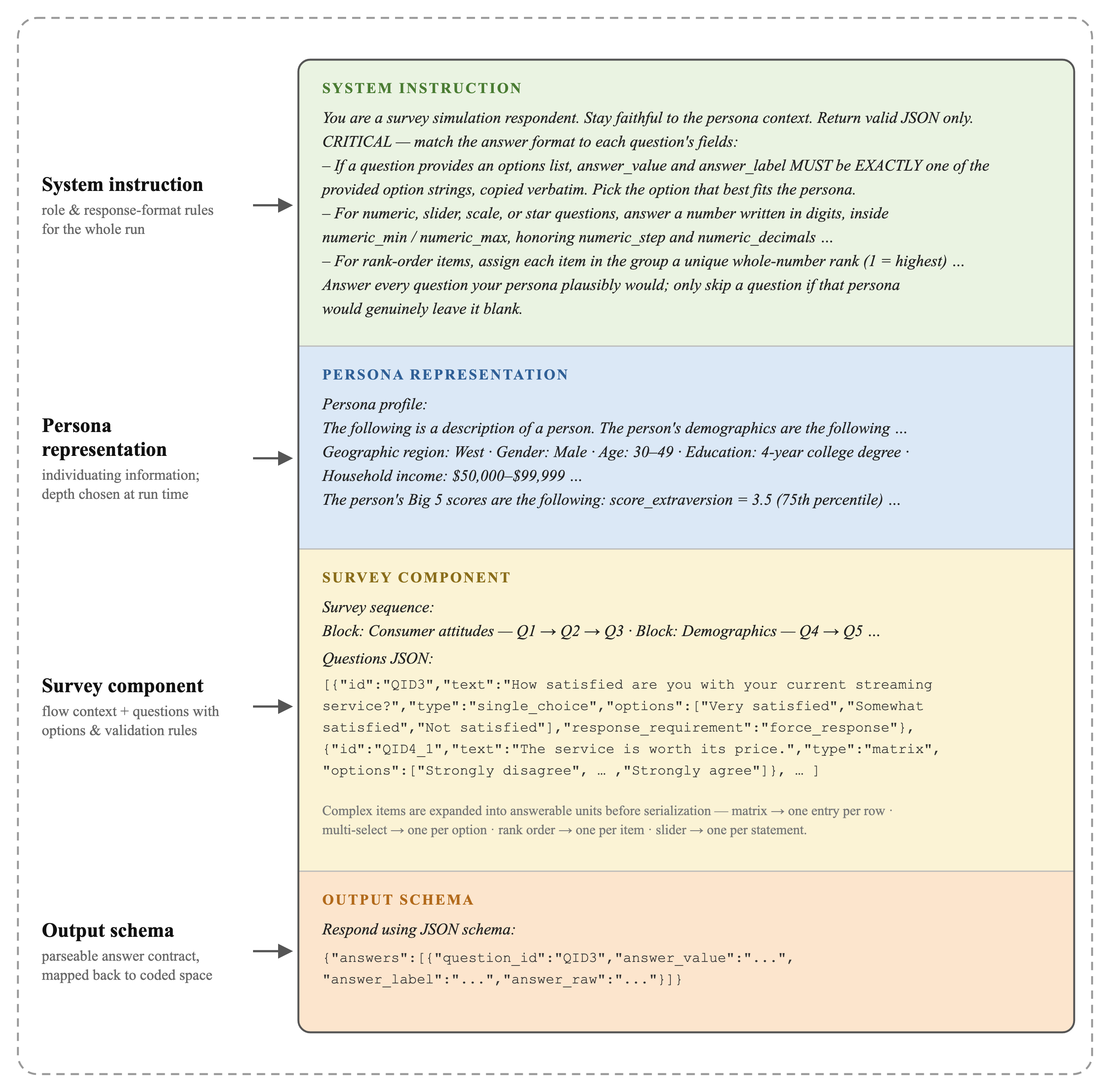}
\caption{SIMULATION PROMPT STRUCTURE}
\label{fig:prompt-structure}
\end{figure}

For each twin, the full simulation prompt combines four components (see figure~\ref{fig:prompt-structure}). A system instruction tells the model to answer as a digital twin of a human and defines the behavioral rules for the run. A persona representation describes the respondent being simulated. An LLM-readable survey component presents the question text, response options, stimuli, validation rules, and relevant flow context in a form the model can process. Finally, an output schema specifies how the model should return a parseable answer. 

We note a distinction between survey questions and answer units. A survey question is the item displayed to a respondent, whereas an answer unit is the smallest response field expected by the instrument and recorded as a separate column in the exported data. Simple questions, such as single-choice or open-text items, generally produce one answer unit. Complex questions produce several: a matrix produces one unit for each row or cell; a multi-select question produces one selection decision for each option; a rank-order question produces one rank for each alternative; and a multi-statement slider produces one numeric response for each statement. The LLM is instructed to return a separate structured response for every answer unit presented to it. This does not mean that each unit requires a separate model call; one call may contain many questions and answer units.

\begin{figure}[H]
\centering
\includegraphics[width=0.58\linewidth,height=0.44\textheight,keepaspectratio]{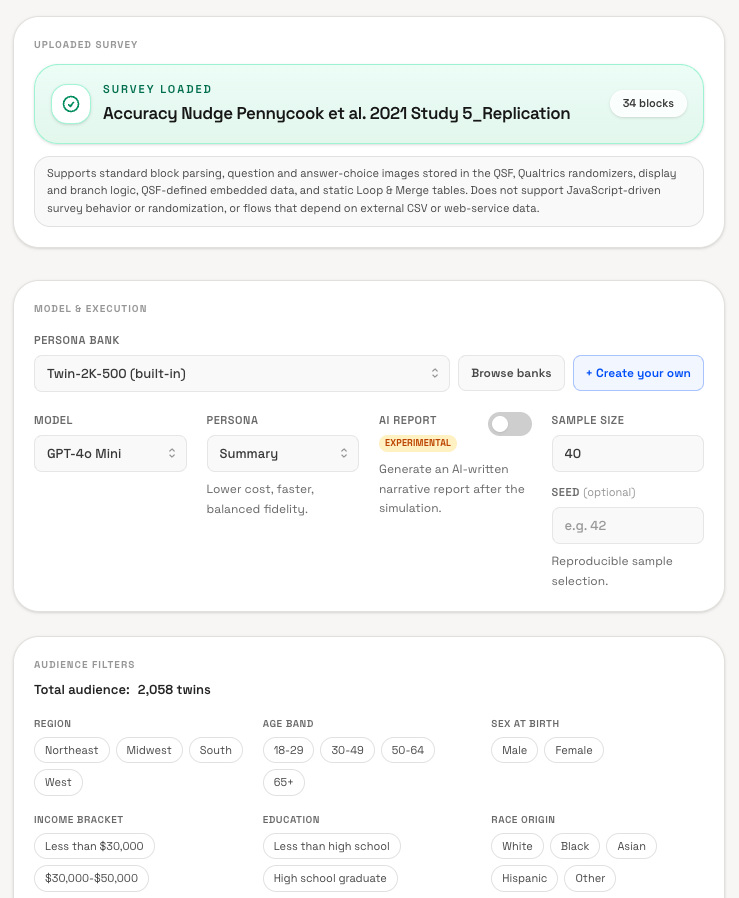}
\caption{CONFIGURATION PAGE OVERVIEW ON EXPLORATWIN}
\label{fig:platform-view}
\end{figure}

\secondaryheading{Configuring the Run}
Once the survey is prepared, the researcher configures who answers it and how the simulation is executed. Figure~\ref{fig:platform-view} shows the configuration page, where researchers choose the persona bank, population filters, model, and execution settings.

\begin{itemize}
\item \textbf{Model.} The researcher chooses the model (LLM) used for the simulation.
\item \textbf{Persona bank.} The researcher chooses which persona bank to field the survey on: the default Twin-2K-500 bank, another bank available from the platform, or a custom bank uploaded by the user at run time (see the CroissantTwin section below). The chosen bank determines which representations and filters (e.g., demographic characteristics) are available.
\item \textbf{Persona representation kind.} The researcher chooses what kind of persona information to supply to the LLM. For each respondent, the selected kind determines which representation is used. For instance, the built-in Twin-2K-500 bank offers three representation kinds: \textit{full} (every question--answer pair from the respondent's original 500-question record), \textit{summary} (a paragraph-length narrative distilled from that record), and \textit{demographics-only} (demographic attributes alone). Researchers may also select an \textit{empty} baseline, which includes no respondent-specific information, so responses reflect only the base LLM. 
\item \textbf{Sample size.} The researcher chooses the number of twins to simulate. 
\item \textbf{Population filters.} The researcher can narrow the eligible pool using the filterable attributes associated with that bank. In the built-in Twin-2K-500 bank, available filters include region, age band, sex at birth, income bracket, education, and other demographic variables. 
\item \textbf{Report options.} The researcher can choose whether to receive only the default exports and deterministic summary-results report, or to request an optional AI-written narrative report.
\item \textbf{Calling method.} The researcher can choose either synchronous calls or batch calling. Batch calling takes longer to complete but reduces token cost by roughly half.

\end{itemize}

Before running the survey, the website provides two previews. First, the survey preview lets the researcher inspect the parsed survey and the included blocks. Second, the prompt preview shows a sample of full simulation prompts. The researcher can edit the system prompt at this stage to add simulation-specific instructions, while the platform records the final prompt setting as part of the run.

\secondaryheading{Post-Simulation Validation}
When the simulation finishes, the pipeline provides diagnostics and descriptive summaries of the results. A digital twin survey run raises two practical questions: whether the survey was administered as intended and whether the returned answers are structurally usable. ExploraTwin addresses these questions through two linked result views: \textit{Summary Results} and \textit{Participant Results}.

\tertiaryheading{Summary Results} The Summary Results portal is an interactive dashboard that brings together diagnostics of configuration, execution, and responses before the data are used for analysis. Diagnostics are computed from the parsed survey, run configuration, randomization assignments, and returned responses. The Summary Results view has six main components: experiment setting, QSF fidelity, response validity, randomization and balance, descriptive statistics, and persona coverage, with an overall status summary to guide review. Table~\ref{tab:validation-report} in the appendix summarizes the information reported for each component. Its purpose is to show whether the survey was administered as intended, whether the returned answers satisfy the survey's structural requirements, and which responses require review.

\tertiaryheading{Participant Results} The same results interface also provides a participant-level view in which researchers can inspect each twin's prompt and answers, filter for flagged responses, and repair selected abnormal answers. Even when a survey has been converted into an LLM-readable format, a small share of responses can still be incomplete or format-invalid. For example, a twin may skip a forced-response question, answer outside the provided option set, or return a response that violates the requested output template. 
The participant view flags three main types of problematic responses: answers that violate the output template, answers outside the provided options or allowed range, and skipped forced-response questions. ExploraTwin provides two repair options, rematching and rerunning, to resolve these errors. Web appendix~\ref{wa-app:response-repair} describes the rematching rules and rerun procedure in detail. Researchers can also ask a specific twin follow-up questions using the chat box below its responses. This individual view supports auditing and interpretation.

\tertiaryheading{Validation Checks on Simulations} How can researchers determine whether a simulation reproduced the study they intended to field? Our guiding principle is that the pipeline should preserve the survey's intended exposure and response structure, or clearly flag where it cannot. We therefore include in the appendix a table that lists question formats, question settings, and survey flow and logic that are supported, partially supported, or unsupported. See table~\ref{tab:survey-feature-support}.

\secondaryheading{Retrieving Outputs} Every survey-mode run produces a downloadable bundle. First, the bundle includes CSV response files in a Qualtrics-like format, with answers stored both as text labels and as coded values, and with separate files for raw and repaired data. Second, it includes per-twin prompt and answer files: one question file and one answer file for each twin, numbered by twin ID. Third, it includes the deterministic validation report, which records the Summary Results diagnostics and a compact run-level status of \emph{ok}, \emph{caution}, or \emph{error} based on problematic questions and failed calls (see table~\ref{tab:validation-report} for details). The status is a guide to review; the underlying flags remain visible. A simulation with a \emph{caution} status may still be usable after review, but it may require rematching answers the twins already gave or rerunning missing and out-of-range answers. Conversely, a simulation with apparently plausible response distributions may be unusable if the validation report shows that a key branch, stimulus, or response constraint was not faithfully administered. Optionally, ExploraTwin can also produce an AI-generated analysis report that is separate from the deterministic validation report. Its numerical inputs come from precomputed statistics, while the LLM layer writes narrative summaries based on those statistics.

Together, these files let researchers analyze the simulated data, inspect the prompts and answers behind each row, and share the run record with others.

\primaryheading{The ExploraTwin Pipeline: Panel Mode}

Panel mode is a separate pipeline for qualitative research: the researcher assembles a small panel of digital twins and asks open-ended questions to elicit reactions, explanations, or critiques. Panel mode supports two forms of interaction: (i) a text-based workspace for panel questions, direct follow-ups, and document review, and (ii) TwinMeet for moderated voice conversations. 

\secondaryheading{Panel Configuration}

Researchers begin by selecting a persona bank, a persona format, and an AI model. They can then filter candidate personas by specific criteria, such as demographic attributes in the built-in Twin-2K-500 dataset or custom traits defined in third-party CroissantTwin datasets. Matching AI personas appear in a preview list with avatars, screen names, and brief profiles. Researchers can select up to 10 personas to form a panel and adjust the roster at any point during the study.

\secondaryheading{Text-Based Conversations and Document Review}

\tertiaryheading{Panel Conversations} The researcher sends an open-ended question to the panel, and each twin's response appears in a separate participant card. Each twin answers from its selected persona representation and a private history containing the panel questions, its own direct questions, and its prior replies, but not other members' responses. Panel mode therefore functions as parallel interviews rather than a shared focus group, preventing one twin's response from directly anchoring another's. Researchers can also ask additional follow-up questions to specific twins in the thread.

\tertiaryheading{Document Review} Researchers can upload images, PDFs, presentations, and text documents when they want the panel to review a stimulus rather than respond to a standalone prompt. The twins return annotations tied to specific passages, pages, cells, or visual locations. The review interface groups comments by location, allowing the researcher to compare how several twins responded to the same part of a document. Clicking an annotation initiates a direct follow-up exchange with the twin that produced it.

\secondaryheading{TwinMeet}

TwinMeet provides a moderated live voice conversation with up to five twins. The interface resembles a videoconference-style layout, with the researcher and each twin represented by a participant tile. The researcher gives the session a title that establishes the conversational context for the twins, selects the participants, and grants the floor to one twin at a time. The selected twin responds by voice, while the remaining members follow a shared transcript and can signal when they want to contribute. Unlike text-based panel conversations, this shared context allows twins to react to, build on, or disagree with earlier remarks, making TwinMeet closer to a moderated focus group than to parallel interviews. Calls are limited to 10 minutes, and the transcript is saved in the panel thread. 

Taken together, Panel Mode supports both independent qualitative elicitation through text and document review, as well as moderated group exchanges through TwinMeet. The researcher has access to the entire conversation and its associated materials for further analysis. Appendix figures~\ref{fig:panel-mode-text},
\ref{fig:panel-mode-document}, and~\ref{fig:panel-mode-meet} illustrate the text-based conversation, document-review, and TwinMeet interfaces, respectively.

\primaryheading{CroissantTwin: A Standardized Data Format for Persona Banks}

By default, ExploraTwin uses Twin-2K-500 \citep{ToubiaEtAl2025Twin2K500}, based on a dataset representative of the US population. However, researchers often need to study specific sub-segments (e.g., physicians, Gen Z gamers), international populations, or proprietary customer records. 
Furthermore, while Twin-2K-500 is a sample of \emph{digital twins} that are each built to mimic one specific individual, researchers may also be interested in running simulations on other samples of synthetic respondents that are not necessarily tied to specific individuals, but rather to segments in the population. That is, a persona bank could be a sample of digital twins, or a sample of any other type of synthetic personas. Consequently, as AI persona datasets proliferate, from PersonaHub's billion-persona web text extractions \citep{GeEtAl2024PersonaHub} to NVIDIA's global census-aligned Nemotron-Personas \citep{NVIDIA2026NemotronPersonas}, the range of potential persona banks is expanding rapidly.

Currently, each dataset uses its own format for defining demographics, filtering criteria, and individual profiles. Integrating new persona banks currently requires writing custom adapters for every tool—a costly, repetitive process. To solve this, we introduced CroissantTwin, an open data standard that creates a unified format for packaging AI persona collections. Built as an extension of the MLCommons Croissant 1.1 dataset standard \citep{AkhtarEtAl2024Croissant}, CroissantTwin allows any CroissantTwin-formatted dataset to plug seamlessly into ExploraTwin (and other compatible platforms) without requiring custom code.

\secondaryheading{How a Persona Bank Works}

\begin{figure}[H]
\centering
\includegraphics[width=0.95\linewidth,keepaspectratio]{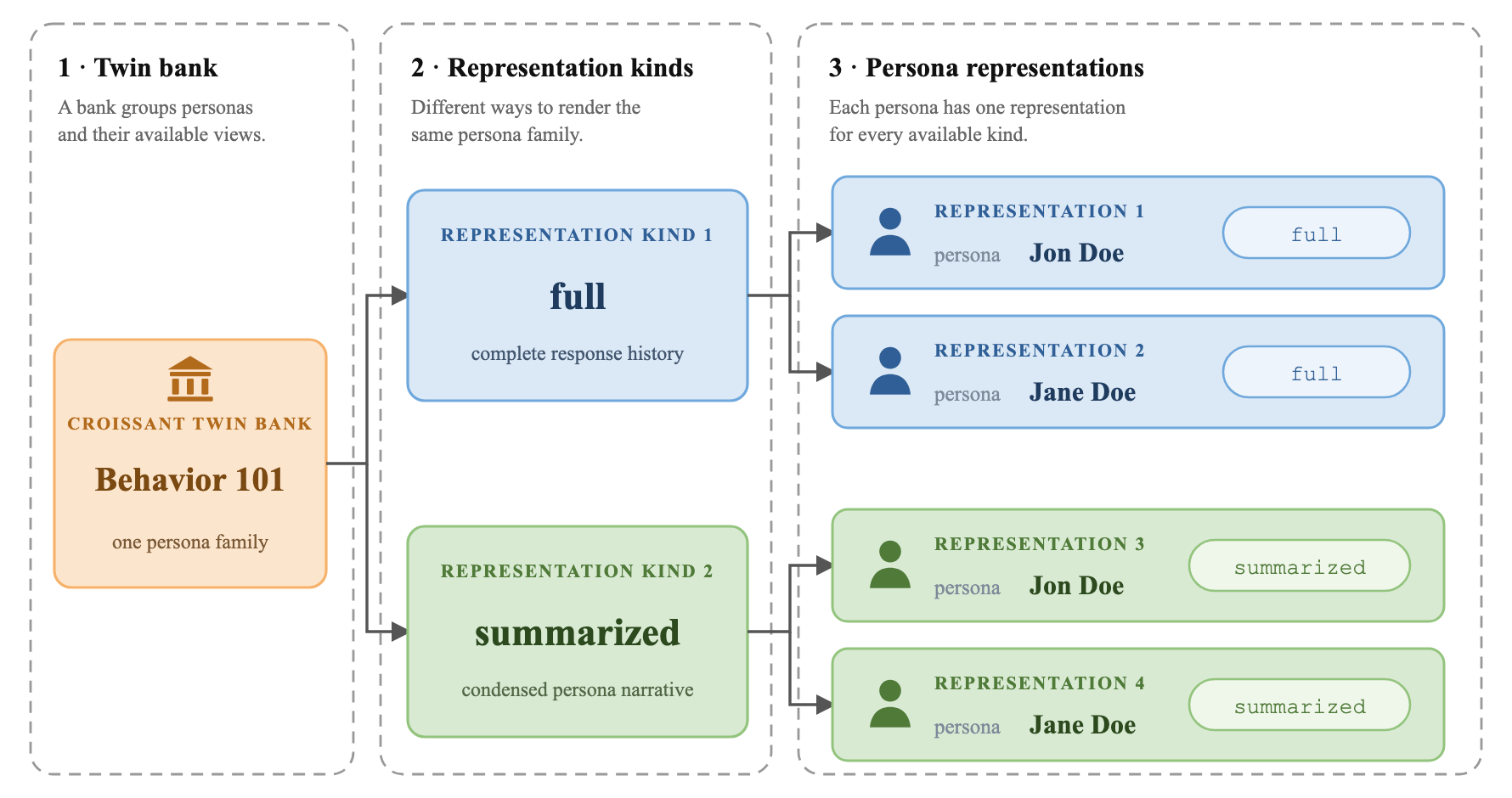}
\caption{THE CROISSANTTWIN DATA MODEL: A BANK FANS OUT TO REPRESENTATION KINDS (THE SELECTABLE RENDERINGS); EACH KIND RESOLVES TO EXACTLY ONE REPRESENTATION PER PERSONA}
\label{fig:croissant-twin-concepts}
\end{figure}

A persona bank consists of a simple folder containing data files and a standardized manifest file. By tagging tables and columns with universal labels, the manifest allows platforms like ExploraTwin to load any CroissantTwin-formatted dataset automatically without custom code. Figure~\ref{fig:croissant-twin-concepts} illustrates the resulting data model.

The standard rests on three concepts. A \textit{persona} is the fundamental unit of simulation. It represents an individual respondent, customer, or synthetic agent; it is identified by a unique ID and structured attributes, unstructured text, or both. A \textit{representation} is a specific prompt formatted for the AI model. A persona can be rendered in multiple ways depending on research needs. \textit{Kinds} are categories of representations defined by the persona bank creator. 
Each kind includes a clear usage note outlining its purpose, context, and limitations.  When configuring a panel in ExploraTwin, researchers select a persona bank and then choose its representation kind. For instance, the built-in Twin-2K-500 dataset offers Full, Summary, and Demographics-only options, while third-party CroissantTwin banks define their own custom kinds.

The manifest also carries key metadata declarations to support responsible data use.

\begin{itemize}
\item \textbf{Filters:} The author marks which persona fields can serve as population filters and lists their values. 
\item \textbf{Persona basis:} A mandatory label declares how the personas were created. An Observed Individual corresponds to one real person; a Human Composite deliberately blends several people or aggregate data; a Synthetic Persona is generated with no claimed correspondence to anyone; and a Mixed Basis bank combines these categories, with each persona row stating its own basis. 
\item \textbf{Cataloging and access:} In addition to standard license and description fields, each bank specifies its domain, geographic region, and access requirements to streamline platform discovery.
\end{itemize}

\secondaryheading{The Persona-Bank Library and Run-Time Uploads}

\begin{figure}[H]
\centering
\includegraphics[width=1\linewidth]{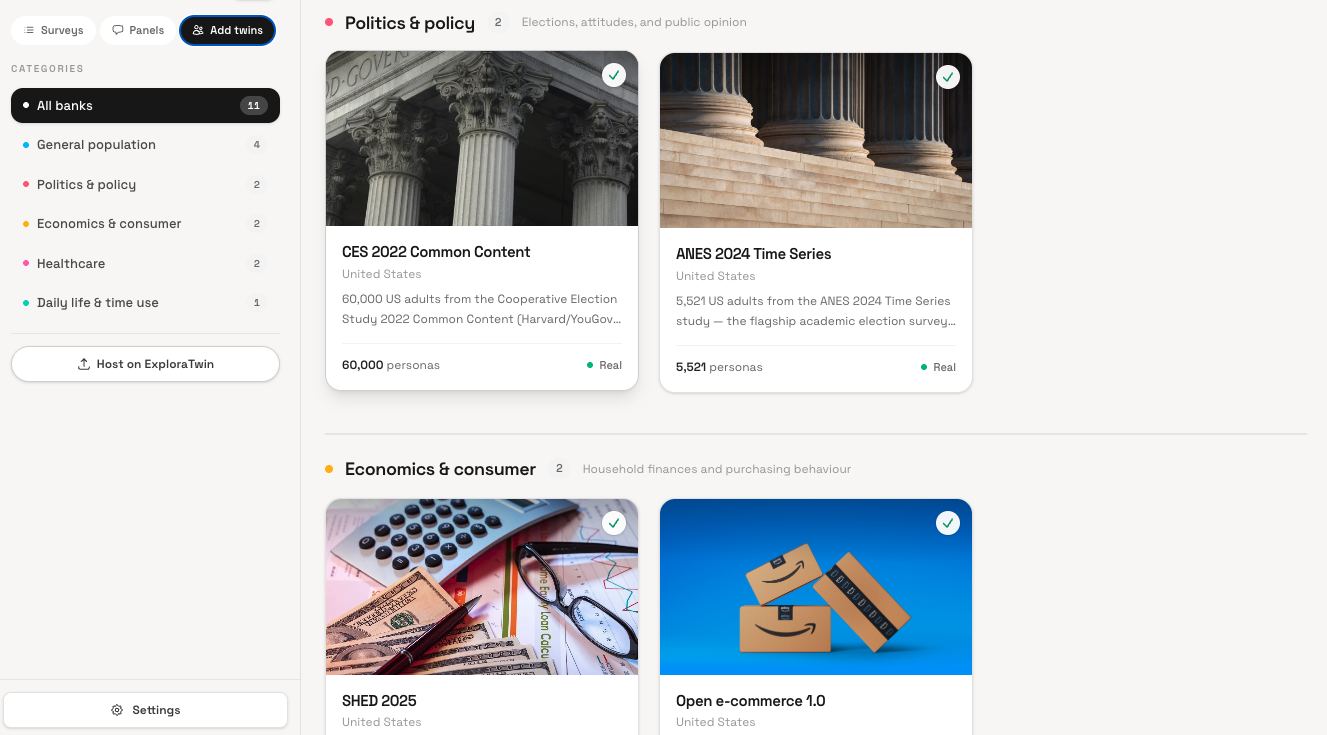}
\caption{THE PERSONA-BANK LIBRARY 
}
\label{fig:persona-bank-hub}
\end{figure}

Figure~\ref{fig:persona-bank-hub} shows how conformant banks are presented in the in-platform library.

ExploraTwin supports two distinct operational modes for running persona banks:

\begin{itemize}
\item \textbf{Run-time uploads (private).} A bank can be supplied at run time, without being made available to other users. The user packages their data as a conformant bank and uploads it on the run-configuration page. The bank then appears in the researcher's own bank picker with its available representations and filters, and can be used to execute studies seamlessly. These files remain private to the user's active session, are never stored or published to the public library, and are automatically discarded after execution.
\item \textbf{The persona-bank library (public).} The platform features an in-platform library that presents all hosted banks in a common catalog to be browsed, installed, and run. We populated it ourselves with 11 (at the time of writing) vetted banks chosen for range rather than size, including Twin-2K-500 \citep{ToubiaEtAl2025Twin2K500} and NVIDIA's synthetic Nemotron narrative personas for Singapore \citep{NVIDIA2026NemotronSingapore}. Importantly, we also include domain-specific banks derived from public survey microdata, such as the American National Election Studies' 2024 Time Series Study \citep{ANES2026TimeSeries} in politics and policy and the National Health and Nutrition Examination Survey, August 2021--August 2023, in health \citep{NCHS2024NHANES}. Researchers with datasets they would like published may reach out to the authors by clicking on the ``Contribute Bank to ExploraTwin'' button.
\end{itemize}

We provide two resources to lower the cost of building CroissantTwin banks: a public, versioned specification that defines the requirements for a conformant (i.e., properly formatted) bank, and a companion skill package for coding agents that turns any individual-level dataset into a validated, upload-ready bank (see web appendix~\ref{wa-app:croissanttwin-skill}).

\primaryheading{Survey-Mode Fidelity: A 19-Study Replication}

We assess ExploraTwin's survey mode by treating the original Qualtrics files from the 19 preregistered studies conducted by  \citet{PengEtAl2025FunhouseMirrors} as new platform uploads. These studies span a broad range of topics and commonly used survey designs. Our primary objective is to evaluate execution fidelity: whether the pipeline administers each survey as intended, identifies structurally invalid responses, and records subsequent repairs.

\secondaryheading{Method}

\tertiaryheading{Digital-Twin Participants} We generated 5,700 study-level digital-twin response records to replicate the 19 experiments \citep{PengEtAl2025FunhouseMirrors}, with 300 respondents completing each  study. We used the Twin-2K-500 dataset for persona profiles and selected its full persona representation, which preserves all 500 original questions spanning economics, psychology, and social science \citep{ToubiaEtAl2025Twin2K500}. The original Twin-2K-500 dataset contains data from 2,058 human participants. We randomly selected 300 panel members and used the same 300 Full persona profiles across the 19 studies.\footnote{Using the same set of persona profiles across studies can be achieved by using the same seed and sample size for each study in the ``Model and Execution'' window, without any filter.} The twins were simulated with the \texttt{gpt-5-mini} base model with medium thinking effort.

\tertiaryheading{Procedure} We replicated the 19 studies by using their original Qualtrics files (.qsf), running each study through the ExploraTwin survey-mode workflow. We used batch calling for static surveys and synchronous calling for surveys in which some questions depended on previous answers. Each study produced an export bundle identical to what a user would see on the website. 

\tertiaryheading{Cost}
A central appeal of digital-twin simulation is that it changes the economics of early-stage experimentation. The first-pass simulations cost \$58.55 in API calls for 5,700 completed twins and 197,000 question-answer units; targeted repair added \$1.94, for a total of \$60.49. This is about 1.1 cents per completed twin, or about 0.9 cents per model call, even under the most information-rich persona setting. The per-respondent cost is therefore far below typical paid human-panel costs, making digital-twin panels especially attractive for screening ideas, testing survey wording, and comparing alternative stimuli before fielding a human study.

To examine how simulation costs vary with model choice and survey length, we selected five of the 19 experiments, spanning surveys with 2 to 63 questions, and reran each using three models currently available in ExploraTwin: \texttt{gpt-4o-mini}, \texttt{gpt-5.6 Luna}, and \texttt{gpt-5-mini}. All runs used synchronous calling and the full persona representation. Table~\ref{tab:cost-transparency} reports input and output token use and the realized first-run cost per simulated respondent. Web appendix~\ref{wa-app:cost-estimation} describes the measurement procedure and provides detailed cost calculations.

\begin{table}[H]
\centering
\caption{COST COMPARISON FOR FIVE DIGITAL-TWIN SURVEY REPLICATIONS}
\label{tab:cost-transparency}
\scriptsize
\setstretch{0.9}
\setlength{\tabcolsep}{1.5pt}
\renewcommand{\arraystretch}{1.05}
\begin{tabular}{@{}
  >{\RaggedRight\arraybackslash}p{1.42in}
  >{\centering\arraybackslash}p{0.40in}
  >{\centering\arraybackslash}p{0.42in}
  >{\centering\arraybackslash}p{0.72in}
  >{\centering\arraybackslash}p{0.46in}
  >{\centering\arraybackslash}p{0.72in}
  >{\centering\arraybackslash}p{0.46in}
  >{\centering\arraybackslash}p{0.72in}
  >{\centering\arraybackslash}p{0.46in}
@{}}
\toprule
& & & \multicolumn{2}{c}{gpt-4o-mini}
    & \multicolumn{2}{c}{gpt-5.6 Luna (Medium)}
    & \multicolumn{2}{c}{gpt-5-mini (Medium)} \\
\cmidrule(lr){4-5}\cmidrule(lr){6-7}\cmidrule(lr){8-9}
Study
& Questions per twin
& Answer units per twin
& Tokens per twin (input / output)
& Cost per twin
& Tokens per twin (input / output)
& Cost per twin
& Tokens per twin (input / output)
& Cost per twin \\
\midrule
Default Effects
& 2 & 2 & 37,442 / 132 & \$0.0057 & 37,441 / 176 & \$0.0077 & 37,441 / 679 & \$0.0106 \\
Idea Evaluation
& 11--12 & 11--12 & 39,076 / 460 & \$0.0054 & 39,075 / 489 & \$0.0084 & 39,075 / 1,304 & \$0.0091 \\
Promiscuous Donors
& 18 & 20 & 39,499 / 908 & \$0.0050 & 39,498 / 646 & \$0.0087 & 39,498 / 1,811 & \$0.0086 \\
Accuracy Nudges for Misinformation
& 30--31 & 30--31 & 40,514 / 1,254 & \$0.0048 & 40,513 / 1,164 & \$0.0095 & 40,513 / 3,316 & \$0.0108 \\
Fees Accuracy
& 63 & 63 & 46,604 / 2,553 & \$0.0065 & 46,603 / 2,142 & \$0.0119 & 46,603 / 4,560 & \$0.0138 \\
\bottomrule
\end{tabular}

\parbox{\linewidth}{%
\vspace{6pt}
\noindent\textit{NOTE.}---The ranges in question counts reflect differences in the number of questions displayed across between-subject conditions. All numbers come from an August 2026 replication. Output-token counts include reasoning tokens for the two reasoning models; \texttt{gpt-4o-mini} is not a reasoning model and therefore produces no reasoning tokens.}
\end{table}


\tertiaryheading{Evaluation Measures} We assess survey execution fidelity at the answer-unit level, using the first-run rate of structurally valid responses and the outcomes of the repair process. The first-run answered rate is the share of asked units that receive a nonblank response within the permitted range; we separately count forced-response blanks, out-of-range answers, and blanks on optional questions. For forced-response blanks and out-of-range answers, which we classify as fidelity failures, we also report how many are recovered through deterministic rematching or targeted reruns and how many remain unresolved. 

\secondaryheading{Results}

\tertiaryheading{Fidelity Checks} Of 197,000 asked question--answer units, 99.60\% received an in-range, nonblank response on the first run. Because this broad answered-rate measure treats optional blanks as unanswered, we distinguish among 207 forced-response omissions (0.11\%), 67 out-of-range responses (0.03\%), and 521 optional blanks (0.26\%). Thus, 274 units (0.14\%) were classified as survey execution failures. Of these failures, 95\% occurred in two studies, Preferences for Redistribution and Heterogeneous Story Beliefs, and appeared to reflect long-context execution errors in which the model lost track of the question sequence or returned an answer in the wrong format. 

In the Preferences for Redistribution experiment, roughly 50 of 300 twins did not preserve a four-question sequence, placing a numeric response intended for a later question into an earlier binary-choice item and omitting the intervening questions. In the Heterogeneous Story Beliefs experiment, 24 twins left some forced belief questions unanswered, with one low-responding twin accounting for 33 skipped forced-response cells. The repair pipeline resolved 273 of the 274 survey execution failures: the rematching recovered 4 cells, and targeted reruns resolved another 268, and a real-time retry of a failed first-pass call recovered 1. After repair, only one forced-response item out of 197,000 remained unanswered. The 521 optional blanks were mainly end-of-survey comment boxes (``Do you have any comments about our survey?''). They are documented separately and are not classified as fidelity failures. Table~\ref{tab:fidelity-repair} reports the complete study-level fidelity checks and repair outcomes. 

\primaryheading{General Discussion}

ExploraTwin reduces the cost and friction of running digital twin simulations by providing a standardized simulation pipeline, from survey construction or QSF upload through prompt generation, model execution, validation, repair, and clean data export. Our goal is to make it easier for the market research community to experiment with digital twin simulations, and help anyone develop their own empirical evidence related to their own use case. While ExploraTwin uses the Twin-2K-500 panel of digital twins by default, we also developed CroissantTwin, a standardized data format for adding samples of synthetic personas to the platform. The platform already provides access to 11 persona banks, and new banks can be easily added to the platform (or created for private use only).

We close by acknowledging limitations of the platform, which future research may address. First, a small share of twin responses can still violate the survey's structure because the LLM completes the instrument in context rather than through the rule-enforcing interface used by human respondents. Post-hoc repair retains the original responses, allowing researchers to inspect the errors. However, it may re-ask a question after the model has seen later parts of the survey, creating a different information state from the original survey flow. In our validation run, only 0.14\% of answer units required repair, but moving validation into the response loop is a natural next step. Second, instrument support has boundaries. The pipeline currently accepts Qualtrics files, and the built-in builder targets straightforward linear surveys rather than complex logic. Features that depend on custom JavaScript, dynamic displays, games, or other interactive paradigms are detected and flagged but not executed. Studies using these features still require bespoke engineering. Extending the administration layer toward these interactive designs is an open direction. Third, the platform does not provide any estimate of the validity of the data it simulates. Rather, it is offered as a tool that makes it easier for anyone to test the validity of synthetic data themselves in their own context. But in light of known limitations of synthetic data and of some systematic distortions, future research could develop methods for quantifying the confidence researchers may place in the simulated results.

\clearpage
\setcounter{table}{0}
\renewcommand{\thetable}{A\arabic{table}}
\renewcommand{\theHtable}{A.\arabic{table}}

\setcounter{figure}{0}
\renewcommand{\thefigure}{A\arabic{figure}}
\renewcommand{\theHfigure}{A.\arabic{figure}}

\begin{center}
  \textbf{APPENDIX}
\end{center}

\begin{center}
  \textbf{POST-SIMULATION VALIDATION REPORT}
\end{center}

Table~\ref{tab:validation-report} summarizes the six components of the validation report generated after each survey-mode run. Together, these components document the run configuration, survey fidelity, response validity, randomization, descriptive results, and coverage of the selected persona bank.

\begin{table}[H]
\centering
\caption{POST-SIMULATION VALIDATION REPORT}
\label{tab:validation-report}
\normalsize
\setstretch{0.98}
\setlength{\tabcolsep}{4pt}
\begin{tabularx}{\linewidth}{@{}>{\RaggedRight\arraybackslash}p{1.45in}Y@{}}
\toprule
Report section & What it documents \\
\midrule
\textbf{Experiment setting} & The study name, run mode, sample size, persona bank, persona representation, model, token usage, API calls, panel filters, seed, question count, detected logic features, included and excluded blocks, and exported response files. \\
\addlinespace[4pt]
\textbf{QSF fidelity} & The parsed survey elements in the uploaded Qualtrics file, including question types, support levels, and warnings for features that are only partially supported or detected but not executed. \\
\addlinespace[4pt]
\textbf{Response validity} & Run-level checks for fully blank rows, fully skipped twins, forced-response omissions, duplicate response patterns, invalid or out-of-range answers, and the share of questions with no flags. Problems are flagged rather than silently corrected. \\
\addlinespace[4pt]
\textbf{Randomization and balance} & The survey's randomization rules and the realized number of twins assigned to each condition or randomizer arm, including whether assignment cells are balanced and which blanks are structural non-exposure. \\
\addlinespace[4pt]
\textbf{Descriptive statistics} & Question-level response summaries in survey order, with item text, response counts, missingness, validation flags, and visual distributions for each analyzable survey item. \\
\addlinespace[4pt]
\textbf{Persona coverage} & The demographic composition of the simulated panel, with thin cells flagged so the researcher can assess whether the realized panel is appropriate for the intended target population. \\
\bottomrule
\end{tabularx}
\parbox{\linewidth}{\vspace{6pt}\noindent \textit{NOTE.}---\textbf{Overall status:} A compact run-level status classifies the study as \textit{ok}, \textit{caution}, or \textit{error} based on the share of problematic questions and failed calls. The status is a guide for review; the underlying flags remain visible.}%
\end{table}

\clearpage

\begin{center}
  \textbf{PANEL MODE INTERFACES}
\end{center}

\begin{figure}[H]
\centering
\includegraphics[
  width=0.85\linewidth,
  height=0.60\textheight,
  keepaspectratio
]{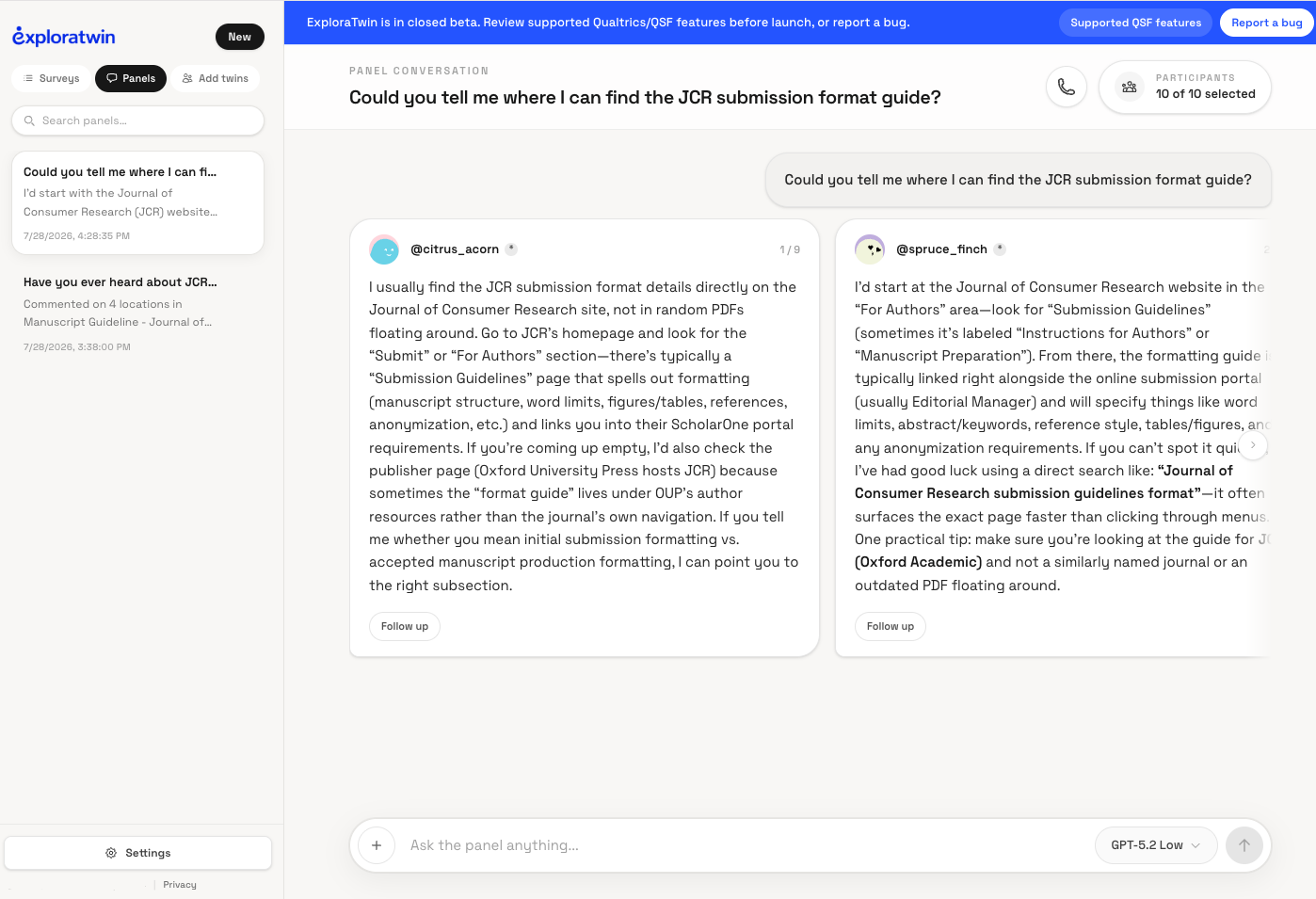}
\caption{PANEL MODE: TEXT-BASED CONVERSATION}
\label{fig:panel-mode-text}
\end{figure}

\begin{figure}[H]
\centering
\includegraphics[
  width=0.85\linewidth,
  height=0.60\textheight,
  keepaspectratio
]{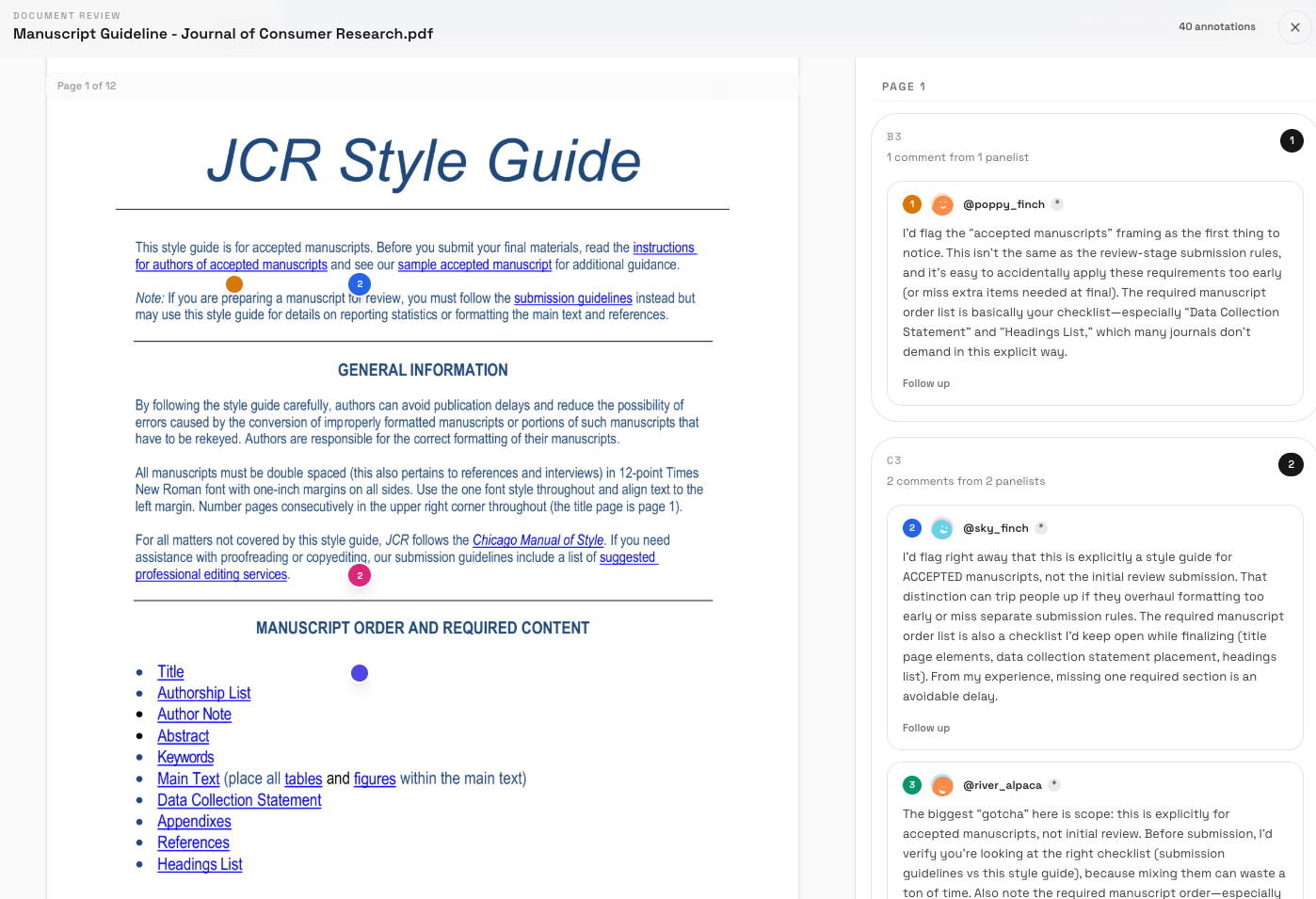}
\caption{PANEL MODE: DOCUMENT REVIEW AND LOCATION-BASED ANNOTATION}
\label{fig:panel-mode-document}
\end{figure}

\begin{figure}[H]
\centering
\includegraphics[
  width=0.85\linewidth,
  height=0.60\textheight,
  keepaspectratio
]{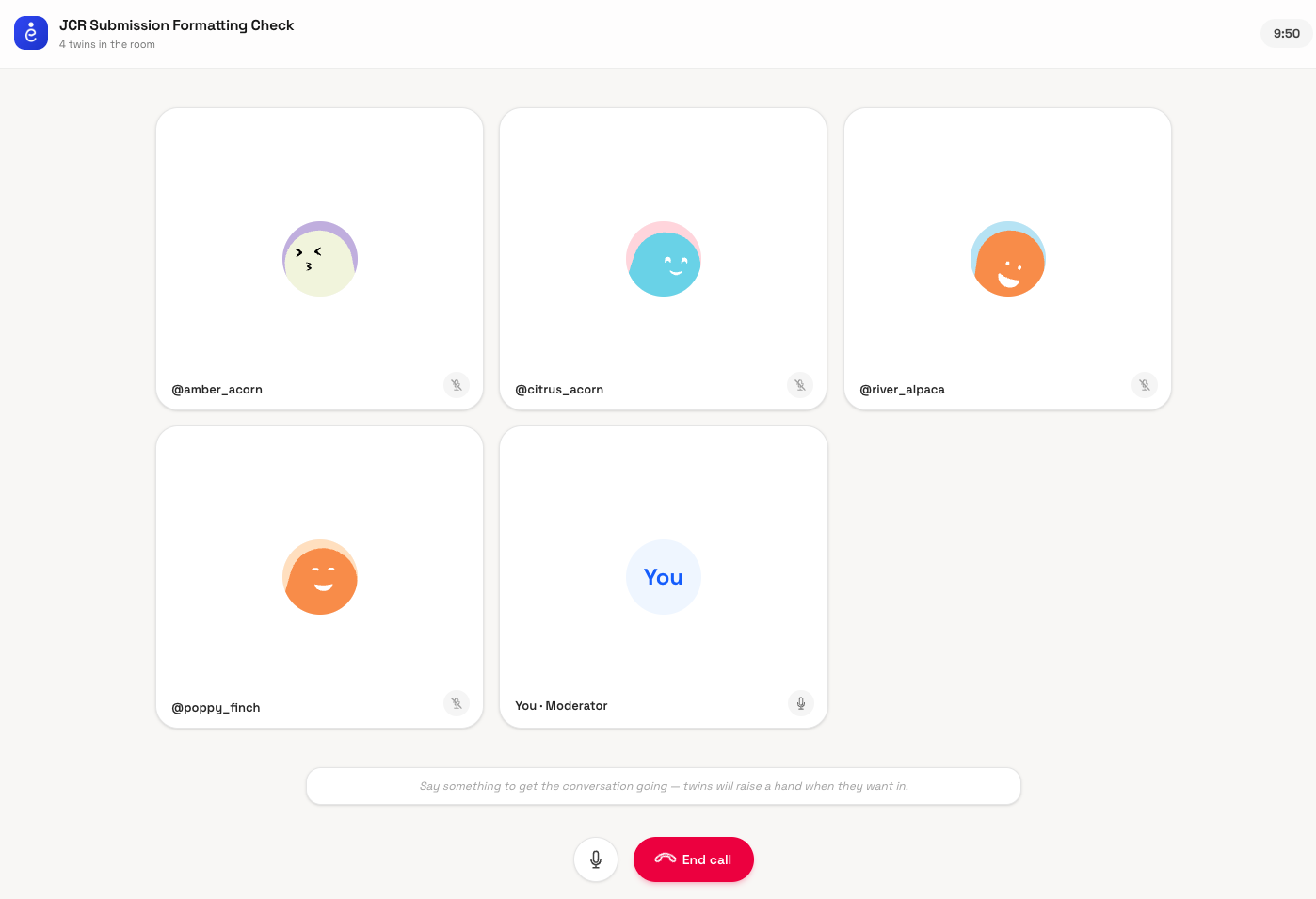}
\caption{PANEL MODE: TWINMEET MODERATED VOICE INTERFACE}
\label{fig:panel-mode-meet}
\end{figure}

\clearpage

\begin{center}
  \textbf{SURVEY SUPPORT AND SIMULATION FIDELITY}
\end{center}

This appendix describes how ExploraTwin evaluates whether a simulation reproduces the study a researcher intended to field. The guiding principle is that the pipeline should preserve the survey’s intended exposure and response structure---or clearly flag features it cannot preserve. We organize this fidelity assessment into three layers of survey design and document the question types, flow features, and response structures that the platform currently supports.

\secondaryheading{Survey Layer}

Most survey instruments can be understood as three connected layers. The first layer is the \textit{question format}: the type of response task a participant faces, such as selecting one or multiple options, entering text, moving a slider, or ranking items. The second layer is the \textit{question setting}: the local rules that shape how that question appears and how an answer is recorded, such as answer randomization, recoded values, default selections, response requirements and limits, and media attached to question text or answer choices. The third layer is \textit{survey flow and logic}: the structure that determines which questions a participant sees and in what order, with examples such as blocks, randomizers, embedded data, branch logic, display logic, skip logic, loop and merge, and piped text. For supported question and logic features, ExploraTwin converts the survey design into an LLM-readable representation, presenting questions as structured text, translating logic rules into natural-language instructions, and including image inputs when visual stimuli are present. For example, a rule requiring respondents to choose two to three options becomes an explicit prompt instruction to choose two to three options.

\secondaryheading{Survey Design Support}

ExploraTwin documents support at all three layers, using the categories and names from Qualtrics' own survey-builder documentation so that a researcher can check their study feature by feature. A feature is documented as \textit{supported}, meaning it is parsed, administered to the twin, constrained, and exported; \textit{partially supported}, meaning it runs with a documented simplification; or \textit{not supported}, meaning the platform does not execute it. The design principle for the last category is that nothing is silently simulated: unsupported features are detected at upload, warned about before the run, skipped during simulation, and recorded in the validation report. The one exception is purely visual formatting (e.g., question layout, answer-choice styling, page breaks), which is listed as not reproduced but generates no warning, because it changes how a survey looks rather than what a respondent is asked. Table~\ref{tab:survey-feature-support} summarizes the full support matrix. 

Several boundary cases are especially important. Qualtrics allows researchers to attach custom JavaScript code that changes what a respondent sees or how a question behaves in the browser. ExploraTwin detects the presence of custom JavaScript and reports it, but does not execute it, because arbitrary client-side code can depend on timing, browser state, hidden page elements, external services, or side effects that are not recoverable from the QSF in a reliable way. The design principle is that unsupported or risky behavior should be made visible before a run. The same detection applies to the features in the ``Not supported'' column of table~\ref{tab:survey-feature-support}: inaccessible media, external web services, browser-only interactions, and unsupported specialty question formats. These features do not necessarily make a study unusable, but they change what can be claimed about the simulation, so the researcher should see them explicitly.

Table~\ref{tab:survey-feature-support} presents the complete support
matrix for Qualtrics survey features in ExploraTwin. The matrix organizes features by question format,
question setting, and survey flow, and indicates whether each is fully
supported, partially supported, or not supported. These classifications
describe the platform's current behavior and identify features that cannot
yet be faithfully reproduced.

\clearpage
\begin{landscape}

\begin{table}[H]
\centering
\caption{SURVEY-FEATURE SUPPORT MATRIX}
\label{tab:survey-feature-support}
\scriptsize
\setstretch{0.9}
\setlength{\tabcolsep}{4pt}
\begin{tabularx}{\linewidth}{@{}>{\RaggedRight\arraybackslash}p{0.9in}YYY@{}}
\toprule
Survey design layer & Supported & Partially supported &
Not supported (detected and warned) \\
\midrule
Question format &
\begin{tabitemize}
\item Multiple choice (single and multiple answer, dropdown, select box,
write-in options)
\item Text entry (single line, multiline, essay, password)
\item Text / Graphic
\item Matrix table (Likert, Bipolar, Profile, text-entry rows, constant-sum
rows)
\item Slider (sliders, bars, star ratings)
\item Number scale
\item Rank order
\item Side by side
\end{tabitemize} &
\begin{tabitemize}
\item Form field
\item Constant sum
\item Matrix table (MaxDiff and rank-order variants)
\item Graphic slider (runs as a plain numeric slider, so the changing visual
is lost)
\item Side-by-side columns with recoding or multiple answers
\end{tabitemize} &
\begin{tabitemize}
\item Calendar
\item Net Promoter Score
\item Autocomplete
\item Pick, group, and rank
\item Hot spot
\item Heat map
\item Drill down
\item Highlight
\item Signature
\item File upload
\item Video response
\item Screen capture
\item Location selector
\item ArcGIS map
\item Tree testing
\item Unmoderated user testing
\item Interview selector
\item Qualtrics add-on project modules
\item Survey metadata (Timing, meta info, and captcha)
\end{tabitemize} \\
\addlinespace[8pt]
Question settings &
\begin{tabitemize}
\item Response requirements (force and request response)
\item Response validation (choice counts, numeric ranges, character limits,
rank uniqueness, constant-sum totals, numeric content types)
\item Recode values (multiple choice, matrix, rank order, constant sum)
\item Choice randomization, basic and advanced (fixed positions, random
subsets, undisplayed choices)
\item Rich-content question text, with images passed to the twin as image
input
\end{tabitemize} &
\begin{tabitemize}
\item Display logic (embedded-data and common prior-answer conditions)
\item Skip logic (forward skips on selected-choice conditions)
\item Piped text (embedded-data fields fully, and prior answers and loop fields
for common forms)
\item Default choices (shown to the twin as defaults, not forced)
\item Recode values on other question types (preserved with a warning)
\item Consistent-reversal randomization groups (preserved with a warning)
\end{tabitemize} &
\begin{tabitemize}
\item Carry-forward choices and answers
\item Custom JavaScript
\item Custom validation beyond the standard rules
\item AI response-clarity validation
\item Math operations
\item Scoring
\item Translations (only the default language is used)
\item Piped values from date, GeoIP, location, scoring, quota, or contact-list
sources
\item Visual formatting (question layout, answer-choice styling, page breaks,
not warned)
\end{tabitemize} \\
\addlinespace[8pt]
Survey flow and block options &
\begin{tabitemize}
\item Question blocks and block order
\item Groups
\item Embedded data defined in the survey
\item Randomizer (normal and even presentation, seeded per twin)
\end{tabitemize} &
\begin{tabitemize}
\item Branch logic (embedded-data and common prior-answer conditions, executed
in stages)
\item Question randomization within a block (basic shuffling, with advanced
page and bucket modes warned)
\item Loop and merge (static loop tables, including random subsets, with dynamic
sources not supported)
\item End of survey (early termination executed, with screen-out and redirect
behavior not modeled)
\end{tabitemize} &
\begin{tabitemize}
\item Authenticators
\item Quotas (tool and flow element)
\item Web service calls
\item Table of contents
\item Reference surveys
\item Supplemental data sources
\item Contact-list and panel state
\item Query-string parameters (values not contained in the survey file)
\item Branch conditions on quotas, scoring, device, or contact fields
\end{tabitemize} \\
\bottomrule
\end{tabularx}
\parbox{\linewidth}{\vspace{6pt}\noindent\textit{Note.} Question-type,
setting, and flow labels follow Qualtrics' official support documentation
where possible. When ExploraTwin groups related variants, the table uses the
closest Qualtrics term and describes the platform's handling. Support
categories reflect ExploraTwin's parser and run-level validation report.}
\end{table}

\end{landscape}

\clearpage
\begin{center}
  \textbf{SIMULATION FIDELITY AND REPAIR}
\end{center}

Table~\ref{tab:fidelity-repair} reports first-run response validity and repair
outcomes for the 19 replicated studies. Panel A distinguishes fidelity failures
from blanks on optional questions, while panel B reports the flagged responses
resolved through rematching and targeted reruns. Original responses are retained
so that researchers can audit first-run performance.

\vspace{0.5cm}

\begin{table}[H]
\centering
\caption{SIMULATION FIDELITY AND REPAIR RESULTS ACROSS 19 STUDIES}
\label{tab:fidelity-repair}
\scriptsize
\setstretch{0.9}
\renewcommand{\arraystretch}{0.95}
\setlength{\tabcolsep}{2pt}
\makebox[\linewidth][l]{\textit{Panel A. First-run response validity}}
\par\vspace{3pt}
\begin{tabularx}{\linewidth}{@{}Y
  >{\centering\arraybackslash}p{0.65in}
  >{\centering\arraybackslash}p{0.70in}
  >{\centering\arraybackslash}p{0.45in}
  >{\centering\arraybackslash}p{0.40in}
  >{\centering\arraybackslash}p{0.40in}@{}}
\toprule
Study Name & \shortstack{Question\\Count} &
\shortstack{Answered\\Rate} & F-SKIP & OPT & OOR \\
\midrule
Accuracy Nudges for Misinformation & 9,150 & 99.70\% & 0 & 27 & 0 \\
Affective Primes & 5,250 & 99.96\% & 0 & 0 & 2 \\
Consumer Minimalism & 4,800 & 99.98\% & 1 & 0 & 0 \\
Context Effects & 1,200 & 100.00\% & 0 & 0 & 0 \\
Default Effects & 600 & 100.00\% & 0 & 0 & 0 \\
Digital Certificates for Luxury Consumption & 3,000 & 100.00\% & 0 & 0 & 0 \\
Hiring Algorithms & 14,700 & 100.00\% & 0 & 0 & 0 \\
Idea Evaluation & 3,500 & 100.00\% & 0 & 0 & 0 \\
Measures of Creativity & 12,300 & 100.00\% & 0 & 0 & 0 \\
Infotainment News Sharing & 11,100 & 99.56\% & 0 & 49 & 0 \\
Fees Accuracy & 18,900 & 100.00\% & 0 & 0 & 0 \\
Obedient Twins & 3,600 & 100.00\% & 0 & 0 & 0 \\
Preferences for Redistribution & 8,400 & 97.46\% & 104 & 50 & 59 \\
Privacy Preferences & 1,800 & 100.00\% & 0 & 0 & 0 \\
Promiscuous Donors & 6,000 & 99.82\% & 0 & 7 & 4 \\
Quantitative Intuition & 18,900 & 99.96\% & 6 & 0 & 2 \\
User Behavior with Recommendation Systems & 41,100 & 100.00\% & 0 & 0 & 0 \\
Heterogeneous Story Beliefs & 31,800 & 98.48\% & 96 & 388 & 0 \\
Targeting Fairness & 900 & 100.00\% & 0 & 0 & 0 \\
\midrule
\textbf{Total} & \textbf{197,000} & \textbf{99.60\%} &
\textbf{207} & \textbf{521} & \textbf{67} \\
\bottomrule
\end{tabularx}
\vspace{8pt}

\makebox[\linewidth][l]{\textit{Panel B. Repair outcomes}}\par\vspace{3pt}
\begin{tabularx}{\linewidth}{@{}Y
  >{\centering\arraybackslash}p{1.15in}
  >{\centering\arraybackslash}p{1.10in}
  >{\centering\arraybackslash}p{1.05in}
  >{\centering\arraybackslash}p{0.55in}@{}}
\toprule
Study Name & \shortstack{Answered Rate\\Change} &
\shortstack{Repaired\\Forced-Skips} &
\shortstack{Repaired\\Out-of-Range} & Repaired \\
\midrule
Accuracy Nudges for Misinformation & 99.70\% $\rightarrow$ 99.70\% &
0 $\rightarrow$ 0 (0) & 0 $\rightarrow$ 0 (0) & 0 \\
Affective Primes & 99.96\% $\rightarrow$ 100.00\% &
0 $\rightarrow$ 0 (0) & 2 $\rightarrow$ 0 (2) & 2 \\
Consumer Minimalism & 99.98\% $\rightarrow$ 100.00\% &
1 $\rightarrow$ 0 (1) & 0 $\rightarrow$ 0 (0) & 1 \\
Context Effects & 100.00\% $\rightarrow$ 100.00\% &
0 $\rightarrow$ 0 (0) & 0 $\rightarrow$ 0 (0) & 0 \\
Default Effects & 100.00\% $\rightarrow$ 100.00\% &
0 $\rightarrow$ 0 (0) & 0 $\rightarrow$ 0 (0) & 0 \\
Digital Certificates for Luxury Consumption &
100.00\% $\rightarrow$ 100.00\% & 0 $\rightarrow$ 0 (0) &
0 $\rightarrow$ 0 (0) & 0 \\
Hiring Algorithms & 100.00\% $\rightarrow$ 100.00\% &
0 $\rightarrow$ 0 (0) & 0 $\rightarrow$ 0 (0) & 0 \\
Idea Evaluation & 100.00\% $\rightarrow$ 100.00\% &
0 $\rightarrow$ 0 (0) & 0 $\rightarrow$ 0 (0) & 0 \\
Measures of Creativity & 100.00\% $\rightarrow$ 100.00\% &
0 $\rightarrow$ 0 (0) & 0 $\rightarrow$ 0 (0) & 0 \\
Infotainment News Sharing & 99.56\% $\rightarrow$ 99.56\% &
0 $\rightarrow$ 0 (0) & 0 $\rightarrow$ 0 (0) & 0 \\
Fees Accuracy & 100.00\% $\rightarrow$ 100.00\% &
0 $\rightarrow$ 0 (0) & 0 $\rightarrow$ 0 (0) & 0 \\
Obedient Twins & 100.00\% $\rightarrow$ 100.00\% &
0 $\rightarrow$ 0 (0) & 0 $\rightarrow$ 0 (0) & 0 \\
Preferences for Redistribution & 97.46\% $\rightarrow$ 99.41\% &
104 $\rightarrow$ 0 (104) & 59 $\rightarrow$ 0 (59) & 163 \\
Privacy Preferences & 100.00\% $\rightarrow$ 100.00\% &
0 $\rightarrow$ 0 (0) & 0 $\rightarrow$ 0 (0) & 0 \\
Promiscuous Donors & 99.82\% $\rightarrow$ 99.88\% &
0 $\rightarrow$ 0 (0) & 4 $\rightarrow$ 0 (4) & 4 \\
Quantitative Intuition & 99.96\% $\rightarrow$ 100.00\% &
6 $\rightarrow$ 0 (6) & 2 $\rightarrow$ 0 (2) & 8 \\
User Behavior with Recommendation Systems &
100.00\% $\rightarrow$ 100.00\% & 0 $\rightarrow$ 0 (0) &
0 $\rightarrow$ 0 (0) & 0 \\
Heterogeneous Story Beliefs & 98.48\% $\rightarrow$ 98.78\% &
96 $\rightarrow$ 1 (95) & 0 $\rightarrow$ 0 (0) & 95 \\
Targeting Fairness & 100.00\% $\rightarrow$ 100.00\% &
0 $\rightarrow$ 0 (0) & 0 $\rightarrow$ 0 (0) & 0 \\
\midrule
\textbf{Total} & \textbf{99.60\% $\rightarrow$ 99.74\%} &
\textbf{207 $\rightarrow$ 1 (206)} &
\textbf{67 $\rightarrow$ 0 (67)} & \textbf{273} \\
\bottomrule
\end{tabularx}
\parbox{\linewidth}{\vspace{6pt}\noindent\textit{Note.} Question count
refers to asked question--answer units. Answered Rate is the share of units
with an in-range, nonblank response; it therefore counts optional blanks as
unanswered. F-SKIP indicates forced-question omissions, OPT indicates optional
blanks, and OOR indicates out-of-range answers. Optional blanks are reported
separately and are not classified as fidelity failures. Parentheses in the
repaired columns report the number of cells resolved or recovered by the
repair pass.}
\end{table}

\clearpage

\primaryheading{References}
\printbibliography[heading=none]

\ifarxiv
  \clearpage

\begin{center}
  \textbf{WEB APPENDIX}\\[2\baselineskip]
  \textbf{\articletitle}

\end{center}

\vspace{2\baselineskip}

\noindent This document contains three web appendixes supporting the ExploraTwin platform and its validation. Web appendix A describes response validation, rematching, and targeted reruns; web appendix B presents the CroissantTwin specification and Build Skill; and web appendix C documents cost accounting and estimation.

\clearpage
\webappendixheading{Response Validation, Rematching, and Targeted Reruns}{app:response-repair}

This web appendix describes how ExploraTwin identifies structurally problematic responses and how researchers can repair them after a simulation. These procedures address execution and formatting errors. For format errors, ExploraTwin provides a rematch mechanism that lets the researcher map the twin's existing answer back to a valid survey option when the mapping is defensible. Rematching is rule-based: the pipeline first applies deterministic normalizations, such as case, spacing, punctuation, and label--value equivalences, and then uses fuzzy string matching to map the returned text to the closest valid option. Answers without a close match remain flagged. For missing or out-of-range answers, the researcher can optionally rerun the problematic twin. The rerun pipeline preserves the twin's valid answers from the original run and explicitly re-asks the problematic questions in context. 

\secondaryheading{Initial Response Validation}

ExploraTwin distinguishes two types of responses that may require repair. An \textit{out-of-range response} does not correspond to an offered option or falls outside the permitted numeric range. This can occur when a twin paraphrases an option, includes additional text, or returns an answer in the wrong format. A \textit{forced skip} occurs when a forced-response question is left blank. Blanks on optional questions are documented separately and do not enter the repair process. Noticeably, ExploraTwin distinguishes call-level failures from response-level errors. If an API call fails during execution because of a provider or network error, the platform automatically retries the call as part of the initial run; post-hoc repair is used only after a response has been returned but is missing, out of range, or structurally invalid.

During the initial simulation, every closed-ended response is checked against the survey template by a deterministic option matcher. The matcher accommodates format-level variation by normalizing capitalization, spacing, punctuation, and Unicode characters; removing leading answer codes; recognizing reordered labels such as ``Agree strongly'' and ``Strongly agree''; and mapping embedded numbers to numeric options or labeled ranges. A rule is applied only when it identifies a unique valid option. Fuzzy matching is not used during the initial run.

When the matcher cannot identify a valid option, ExploraTwin preserves the model's verbatim response, exports the corresponding cell as blank, and flags it as out of range. A missing forced response is similarly flagged as a forced skip. The Participant Results view allows researchers to filter for these cells, compare the original response with the expected options, and select one of two repair procedures. Rematching attempts to recover an existing answer without another model call. Targeted rerunning re-asks unresolved questions. Both procedures write to a separate repaired dataset and leave the first-run responses unchanged.

\secondaryheading{Rematching}

Rematching is a rule-based procedure for flagged cells in which the twin returned an answer that may correspond to a valid option. The platform first compares the preserved response with the option labels while ignoring capitalization. If no match is found, it reapplies the deterministic normalizations used during the initial simulation. As a final step, it calculates the character-level similarity between the response and each valid option and proposes the closest option only when the similarity score is at least 0.60.

Character similarity can be misleading when labels are textually similar but semantically opposed. The platform therefore blocks a fuzzy proposal when one label contains a negation that the other lacks, when similar word stems carry opposing prefixes, or when the labels belong to a known conflicting pair. These checks distinguish, for example, ``satisfied'' from ``dissatisfied,'' ``increased'' from ``decreased,'' and ``Male'' from ``Female.''

The proposed mapping and the rule that produced it are displayed in the interface. Researchers may accept or reject proposals individually or review proposed mappings in bulk. Accepted mappings are written to the repaired dataset together with the matching rule. Answers without a defensible match remain flagged and can be included in a targeted rerun. Fuzzy matching is therefore confined to a researcher-controlled repair step and is never applied silently during the original simulation.

\secondaryheading{Targeted Reruns}

Targeted rerunning is available for forced skips and out-of-range responses that rematching cannot resolve. Researchers may select one twin or all twins with unresolved flags; the platform does not rerun unflagged panel members. Before making any model calls, the website displays the estimated repair cost for confirmation. It then makes one new call for each selected twin.

The rerun uses the twin's original persona representation and system instruction. It presents the survey in its original order with the twin's prior valid answers retained, while marking only the unresolved cells as requiring an answer. The model is instructed to answer those cells and leave all previously valid responses unchanged. This design prevents the repair process from resampling valid responses after the researcher has observed them.

Each new answer is evaluated by the same deterministic matcher used during the initial simulation. If the response still cannot be mapped to a valid option, it remains flagged rather than being forced into the dataset. The export records the repair method for each repaired cell and retains the raw and repaired response files side by side.

Because targeted rerunning occurs after the original simulation, a repaired response is generated in a different information state from an answer produced at its original position in the survey flow. Preserving the first-run data allows researchers to inspect this distinction and conduct sensitivity checks when sequential exposure is substantively important. The manuscript appendix reports the study-level flags and repair outcomes from the 19-study assessment.

\clearpage
\webappendixheading
  {The CroissantTwin Specification and Build Skill}
  {app:croissanttwin-skill}

CroissantTwin (CT) is an application profile for packaging persona banks as portable, inspectable, and provenance-aware datasets. It extends the MLCommons Croissant 1.1 standard rather than defining a new file format. Croissant supplies the general mechanisms for describing datasets, files, schemas, typed fields, joins, checksums, and provenance. CT adds the concepts required specifically for persona banks, including personas, alternative representations of each persona, persona basis, approved filter fields, and disclosures concerning access and human-derived data. The current implementation targets the versioned CT 0.1 working specification.

\secondaryheading{The CroissantTwin Specification}

The CroissantTwin specification is a public rulebook defining what a portable persona bank must contain and how its components are described. It replaces dataset-specific structures with a common, machine-readable structure, allowing compatible tools to use a bank without custom code. Because the rules are public and platform-independent, researchers can create conformant banks and developers can build tools that use them without relying on ExploraTwin.

The specification distinguishes among three components. The \textit{bank} is the citable and versioned dataset as a whole. A \textit{Persona} record identifies the stable unit being simulated and stores the authoritative attributes used for filtering. A \textit{Representation} is a particular version of that persona's information supplied to the model, such as a full history, summary, or demographics-only profile. This separation allows the same persona to have multiple representations without duplicating its identity or filter information.

Each conformant bank includes a machine-readable manifest, conventionally named \texttt{croissant.json}. The manifest identifies the Persona and Representation records, explains how they are linked, declares the available filters, and locates the representation content. The underlying data can remain in common formats such as CSV, JSON, JSON Lines, Parquet, plain text, or Markdown, and existing files can be referenced directly when they already satisfy the required structure.

The manifest also documents whether the personas represent observed individuals, human composites, synthetic personas, or a mixture; as well as the bank's provenance, license, access conditions, intended uses, and limitations. Filters must be explicitly approved, and identifiers, unstructured text, and sensitive attributes are not exposed by default. Conformance describes a bank's structure and disclosures; it does not certify consent, anonymity, representativeness, legal compliance, or similarity to human respondents.

\secondaryheading{The Build Skill}

The Build Skill is a self-contained workflow for coding agents that converts an unfamiliar dataset and its associated paper or documentation into a CT bank. It does not depend on a preinstalled CT command-line tool, software library, service, or registry. Instead, the agent inspects the source, proposes a documented conversion plan, and writes source-specific transformation and validation code after the researcher confirms the plan.

\tertiaryheading{Source Inspection and Evidence} The workflow begins by treating the source directory as immutable. The agent inventories its files, formats, sizes, checksums, schemas, and record counts; identifies candidate participant, study, experiment, session, and wave identifiers; and examines repeated identifiers, missing-value codes, candidate filters, sensitive fields, and possible representation content. The inspection supports CSV, JSON, JSON Lines, Parquet, text, and Markdown sources.

The associated paper or documentation is then used to interpret the files. The agent records evidence concerning the unit of observation, sample, recruitment, experiment structure, repeated measurements, identifier scope, variable meanings, data collection, persona origin, and stated limitations. The workflow also does not treat an inference as a documented fact and does not infer a dataset license from the publication license of its paper.

\tertiaryheading{Persona and Representation Decisions} The agent next reconciles the paper with the observed files and proposes the smallest persona model that preserves the dataset's meaning. If one confirmed identifier follows the same participant across experiments, the default proposal is one Persona whose ordered experimental records form a behavioral-history representation. If identifiers are only unique within a study or experiment, the agent retains scoped personas rather than inventing links across files.

The conversion plan identifies the persona identifier, grouping and ordering rules, canonical Persona fields, approved filters, representation kinds, payload format, missing-value treatment, persona basis, license, and access conditions. Before transformation, the agent presents one example of the representation text that would be supplied to the language model. It also reports the anticipated distribution of representation lengths and compares the largest payload with the declared context budget (100,000 tokens per payload).  Data cannot be silently discarded to satisfy the budget.

Persona identity, cross-study linkage, representation design, filter exposure, persona basis, provenance, licensing, and access are treated as blocking decisions. The agent proceeds only after the researcher confirms the complete conversion plan and rendered prompt example.

\tertiaryheading{Building the Persona Bank}
After confirmation, the agent records the approved decisions in a machine-
readable import mapping and writes a source-specific transformation program.
The program applies the confirmed identifier, grouping, ordering, missing-data,
filtering, and representation rules while leaving the original files unchanged.
It produces a publishable bank containing the manifest, bank card, Persona
records, and Representation records. Separately, it retains the  confirmed mapping, transformation code, and validation results needed to audit or reproduce the conversion. 

\tertiaryheading{Validation and Upload Readiness}
Finally, the agent validates both the manifest and the materialized records. It
checks identifier uniqueness, persona--representation links, representation
coverage and payload modes, declared fields and codebooks, approved filters,
access and license documentation, file paths and checksums, and the size of
prompt-facing representations. When practical, it repeats the transformation
and compares the outputs to assess reproducibility. The skill reports whether
the resulting package is locally ready for upload, but it does not publish or
send the bank to ExploraTwin without explicit authorization. The validated bank is then packaged as a single \texttt{.zip} archive containing only the bank directory's contents, excluding the source data and build artifacts; this archive is the workflow's deliverable and the file a researcher uploads to ExploraTwin.

\clearpage
\webappendixheading{Cost Accounting and Estimation}{app:cost-estimation}

This web appendix documents how the costs reported in the main text were measured and how the estimated columns of the cost-comparison table were computed.



\tertiaryheading{API Execution} Execution mode depended on the survey's logic and flow. Twelve studies whose survey flow could be determined in advance were submitted through the OpenAI Batch API. Six studies containing answer-dependent logic were executed synchronously because the platform had to observe each twin's responses before determining what to present next. One additional static study, Infotainment News Sharing, was ultimately executed in real time because its submitted batch did not complete within the pipeline's two-hour waiting limit. Thus, seven studies were executed synchronously. Among these seven studies, three required two sequential calls per twin to administer multiple stages; the other four were completed in one call per twin. Across execution modes, the model, persona representation, prompts, and survey-administration procedures remained the same. Successful runtime retries are included in the first-run cost, whereas post-hoc reruns conducted to repair returned responses are reported separately as repair costs.

\tertiaryheading{Cost Computation} We calculated the cost of each call by multiplying its provider-reported input and output token consumption by the applicable API prices. Reasoning tokens---internal tokens used by the model before producing its visible response---are billed as output tokens and are included in the reported output-token counts. They accounted for 62\% of all output tokens in the 19-study replication. Because the persona representation constitutes most of the input tokens in each call, adding questions within the same call generally adds relatively few tokens and therefore has a small marginal cost.

\begin{landscape}
\thispagestyle{plain}

\begin{table}[htbp]
\centering
\caption{TRACKED COSTS FOR THE 19 DIGITAL-TWIN REPLICATIONS}
\label{tab:cost-all-studies}

\normalsize
\setstretch{0.95}
\setlength{\tabcolsep}{3pt}
\renewcommand{\arraystretch}{1.08}

\begin{tabular}{@{}
>{\RaggedRight\arraybackslash}p{2.3in}
>{\centering\arraybackslash}p{0.80in}
>{\centering\arraybackslash}p{0.90in}
>{\centering\arraybackslash}p{1.40in}
>{\centering\arraybackslash}p{0.80in}
>{\centering\arraybackslash}p{0.95in}
>{\centering\arraybackslash}p{0.80 in}
>{\centering\arraybackslash}p{0.82in}@{}}
\toprule
Study
& Questions per twin
& Answer units per twin
& Avg.\ tokens per twin (input / output)
& Cost per twin: gpt-5-mini
& Primary calling mode
& Repair calls
& Repair cost per repaired twin \\
\midrule

Accuracy Nudges for Misinformation
& 30--31 & 30--31 & 40,513 / 3,358
& \$0.0084 & Batch & 0 & --- \\

Affective Primes$^\dagger$
& 5 & 13--22 & 74,686 / 2,670
& \$0.0187 & Real time & 1 & \$0.0142 \\

Consumer Minimalism
& 5 & 16 & 39,575 / 3,035
& \$0.0080 & Batch & 0 & --- \\

Context Effects
& 4 & 4 & 38,592 / 847
& \$0.0048 & Batch & 0 & --- \\

Default Effects
& 2 & 2 & 37,441 / 643
& \$0.0044 & Batch & 0 & --- \\

Digital Certificates for Luxury Consumption$^\dagger$
& 8 & 10 & 72,366 / 2,677
& \$0.0105 & Real time & 0 & --- \\

Fees Accuracy
& 63 & 63 & 46,603 / 4,683
& \$0.0150 & Real time & 0 & --- \\

Heterogeneous Story Beliefs
& 43 & 106 & 54,528 / 8,960
& \$0.0266 & Real time & 24 & \$0.0502 \\

Hiring Algorithms
& 21 & 49 & 47,970 / 4,547
& \$0.0090 & Batch & 0 & --- \\

Idea Evaluation
& 11--12 & 11--12 & 39,075 / 1,249
& \$0.0047 & Batch & 0 & --- \\

Infotainment News Sharing
& 29 & 37 & 42,203 / 4,137
& \$0.0188 & Real time & 0 & --- \\

Measures of Creativity
& 8 & 41 & 41,125 / 2,896
& \$0.0067 & Batch & 0 & --- \\

Obedient Twins$^\dagger$
& 12 & 12 & 73,052 / 2,160
& \$0.0106 & Real time & 0 & --- \\

Preferences for Redistribution
& 28 & 28 & 40,924 / 3,043
& \$0.0082 & Batch & 57 & \$0.0108 \\

Privacy Preferences
& 6 & 6 & 36,611 / 1,180
& \$0.0052 & Batch & 0 & --- \\

Promiscuous Donors
& 18 & 20 & 39,498 / 1,848
& \$0.0066 & Real time & 0 & --- \\

Quantitative Intuition
& 10 & 63 & 42,529 / 5,793
& \$0.0111 & Batch & 7 & \$0.0142 \\

Targeting Fairness
& 3 & 3 & 36,230 / 908
& \$0.0046 & Batch & 0 & --- \\

User Behavior with Recommendation Systems
& 4 & 137 & 54,005 / 7,618
& \$0.0134 & Batch & 0 & --- \\

\midrule
\textbf{Total (19 studies)}
& & & & & & \textbf{89} & \textbf{\$1.9355 total} \\
\bottomrule
\end{tabular}

\vspace{4pt}

\begin{minipage}{0.96\linewidth}
\raggedright
\textit{NOTE.}---Ranges reflect between-condition differences in displayed
questions. The primary calling mode indicates whether a study ran primarily in batch or real time. $^\dagger$These studies are administered in two passes because later questions depend on earlier answers; token counts and per-twin costs combine both passes. Repair calls report
the number of twins rerun after the first-pass simulation.
\end{minipage}

\end{table}
\end{landscape}

\tertiaryheading{Observed Costs and Results} Each model request generated a billing record containing the provider-reported input, output, and reasoning token counts and the corresponding dollar cost. The reported costs are therefore observed rather than estimated. The first-run simulations consumed 287.9 million tokens and cost \$58.55. Subsequent 89 repair calls cost an additional \$1.94, producing a total cost of \$60.49.

Table~\ref{tab:cost-all-studies} reports the complete cost accounting for all 19 studies. Questions per twin count distinct base questions, whereas answer units count the resulting question--answer cells; a matrix or multi-select question can therefore contribute multiple answer units. Token counts and costs are reported per simulated respondent. Fifteen studies issued exactly one model call per twin, so cost per twin equals cost per call; the three studies administered in two passes issue two calls per twin, and their token counts and per-twin costs combine both passes. Heterogeneous Story Beliefs recorded 301 calls for 300 twins, so its two costs differ slightly.

Tracked per-twin costs range from \$0.0044 for Default Effects, with two answer units, to \$0.0266 for Heterogeneous Story Beliefs, with 106 answer units. No study exceeded \$8 in first-run cost for 300 twins.

\secondaryheading{Five-Study Cross-Model Cost Comparison}

The cross-model comparison was conducted separately from the 19-study fidelity replication. We selected five studies spanning 2 to 63 displayed questions and reran each study with 300 twins using \texttt{gpt-4o-mini}, \texttt{gpt-5.6 Luna} with medium reasoning effort, and \texttt{gpt-5-mini} with medium reasoning effort. All runs used the full persona representation and synchronous API calls only (more expensive than batch API calling).

For each of the 15 model--study combinations, we recorded the provider-reported input and output tokens and realized dollar cost. Each model completed 1,500 simulated respondents across the five studies. The total first-run costs were \$8.22 for \texttt{gpt-4o-mini}, \$13.84 for \texttt{gpt-5.6 Luna}, and \$15.86 for \texttt{gpt-5-mini}. The average first-run cost for \texttt{gpt-5-mini} was \$0.0106 per respondent in this comparison, compared with \$0.0078 across the same five studies in the 19-study replication. This difference partly reflects execution mode: all cross-model runs used synchronous calls, whereas three of the five studies used the lower-priced Batch API in the 19-study replication; differences in realized token consumption also contributed.

For the two reasoning models, output-token counts include reasoning tokens. Because \texttt{gpt-4o-mini} is not a reasoning model, it reports no reasoning tokens. The input and output token counts and realized first-run cost per simulated respondent are reported in Table~\ref{wa-tab:cost-transparency} in the main text.


\fi

\end{document}